\documentclass[english]{article}

\usepackage{graphicx}
\usepackage{url}

\newcommand\ff[1]{#1}

\title{Latitude and power characteristics\\ of solar activity\\ in the end of the Maunder minimum}
\author{V.G.~Ivanov and E.V.~Miletsky\\[1ex]
Central Astronomical Observatory at Pulkovo,\\
Saint-Petersburg, Russia\\[1ex]
E-mail: vg.ivanov@gao.spb.ru
}
\begin{document}

\maketitle

\begin{abstract}
Two important sources of information about sunspots in the
Maunder minimum are the Sp\"orer catalog \cite{spoerer} and observations of the
Paris observatory \cite{ribes}, which cover in total the last quarter of
the 17th and the first two decades of the 18th century.

These data, in particular, contain information about sunspot latitudes.
As we showed in \cite{imn2011, im2016}, dispersions of sunspot latitude
distributions are tightly related to sunspot indices, so we can
estimate the level of solar activity in this epoch by a method
which is not based on direct calculation of sunspots and
is weakly affected by loss of observational data.

The latitude distributions of sunspots in the time of transition
from the Maunder minimum to the common regime of solar activity
proved to be wide enough. It gives evidences in favor of, first, not
very low cycle No.~$-3$ (1712--1723) with the Wolf number in maximum
$\mbox{W}=100\pm50$, and, second, nonzero activity in the maximum of
cycle No.~$-4$ (1700--1711) $\mbox{W}=60\pm45$.

Therefore, the latitude distributions in the end of the Maunder minimum are in
better agreement with the traditional Wolf number
and new revisited indices of activity SN and GN \cite{sn,gn} than with the
GSN \cite{gsn1998}; the latter provide much lower level of activity in
this epoch.
\end{abstract}

\section{Introduction}

The epoch of the Maunder minimum (MM) \cite{eddy1976} lasted, as it
is traditionally believed, since the middle of the 17th to the
beginning of the third decade of the 18th century. It was very
special in a low level of solar activity as well as its eminent
hemisphere asymmetry. Now nobody doubts that the activity of the Sun
in this epoch was low; however, it is still discussed how low it was
(see, e.g., \cite{usoskin2015, zolotova2015}). The related question
is when the solar activity turned back to its normal regime. The
answers of these questions are tangled by the fact that the
observations of sunspots in this epoch are incomplete. That is why
the problem of valid estimations of solar activity indices on the
base of fragmentary observational data is of special importance.

Two important sources of information about sunspot group during the MM
are the Sp\"orer catalog of sunspots \cite{spoerer} and the
observations of the Paris observatory  \cite{ribes}, which in total
cover the most part of the epoch of grand minimum (1672--1719). These sources include
information not only on numbers, but also on heliolatitudes of
sunspots. As we showed in \cite{imn2011,im2016},  there is a high correlation
between the latitude dispersions of sunspots and power of solar activity.
Therefore, we can made independent estimates of the activity level in this epoch.

\begin{figure}
\begin{center}
\ff{\includegraphics[width=0.99\textwidth]{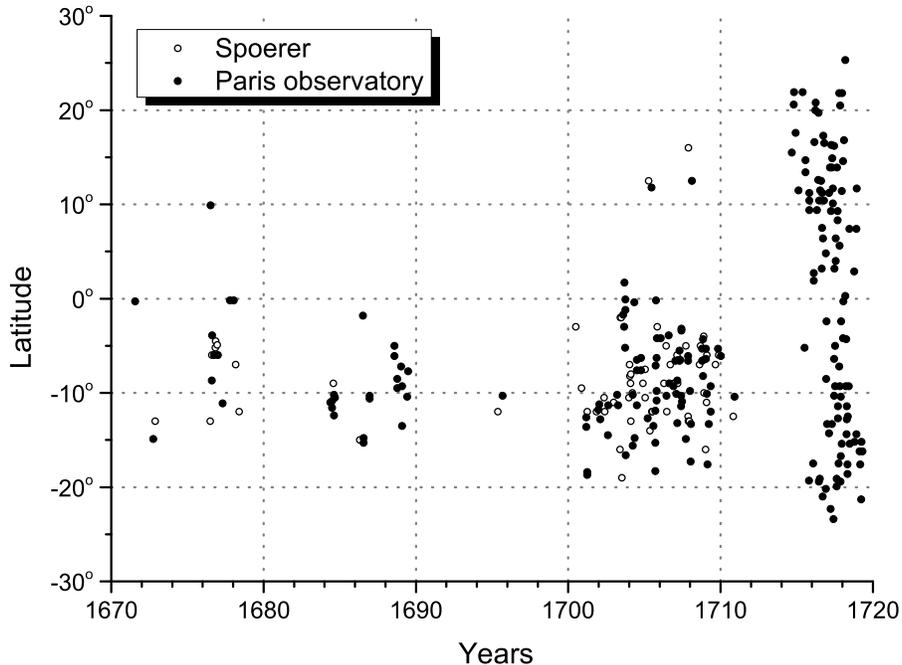}}
\caption{The ``Maunder butterflies'' for the Sp\"orer catalog \cite{spoerer} (empty circles)
and the observations of the Paris observatory\cite{ribes} (filled circles).} \label{fig1}
\end{center}
\end{figure}

\section{Data and method}

We used latitudes of sunspots from the paper of Sp\"orer \cite{spoerer}
(64 observations) and observations of the Paris observatory (213
observations), which were digitized and compiled to a single
catalog in \cite{vaquero2015}. The ``Maunder butterflies'' diagram for these
catalogs are plotted in Fig.~\ref{fig1}. For these data we calculated
the ``index of sunspot groups'' G, which is equal to yearly averaged
number of daily observed groups, and yearly dispersions of
absolute values of heliographic latitudes of sunspots
$\sigma_\phi^2$.

In these data it is usually unclear whether a single sunspot or a
sunspot group was observed, and we will treat all observations as
groups. Treating them in opposite way, i.e. as individual sunspot,
would affect G but not $\sigma_\phi$, and it is the latter values
that are of primary importance for us.

We also calculated indices G and $\sigma_\phi^2$ for
the extended Greenwich/NOAA catalog (GC) (1875--2015) \cite{greenwich}.

\begin{figure}
\begin{center}
\ff{\includegraphics[width=0.55\textwidth]{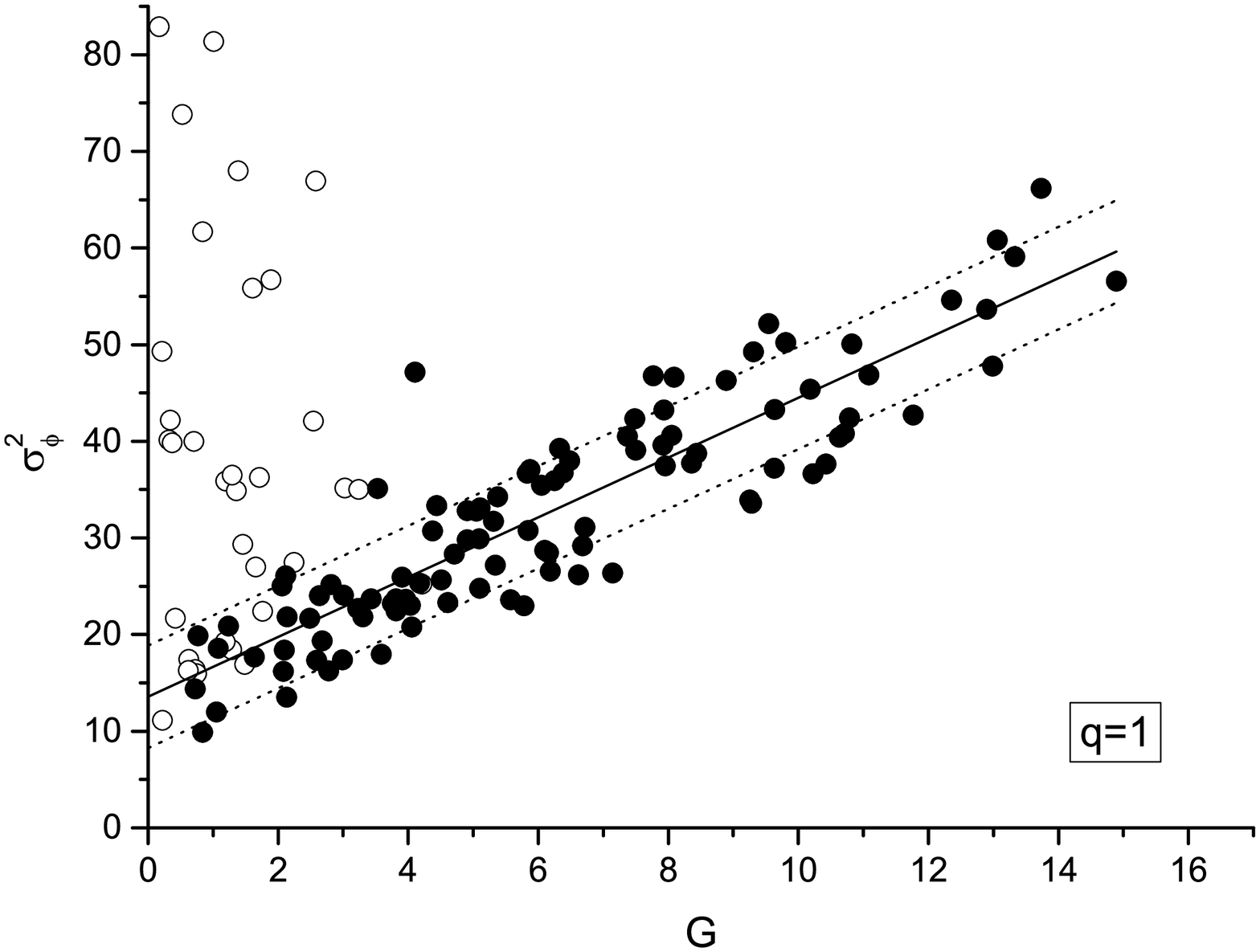}}
\ff{\includegraphics[width=0.55\textwidth]{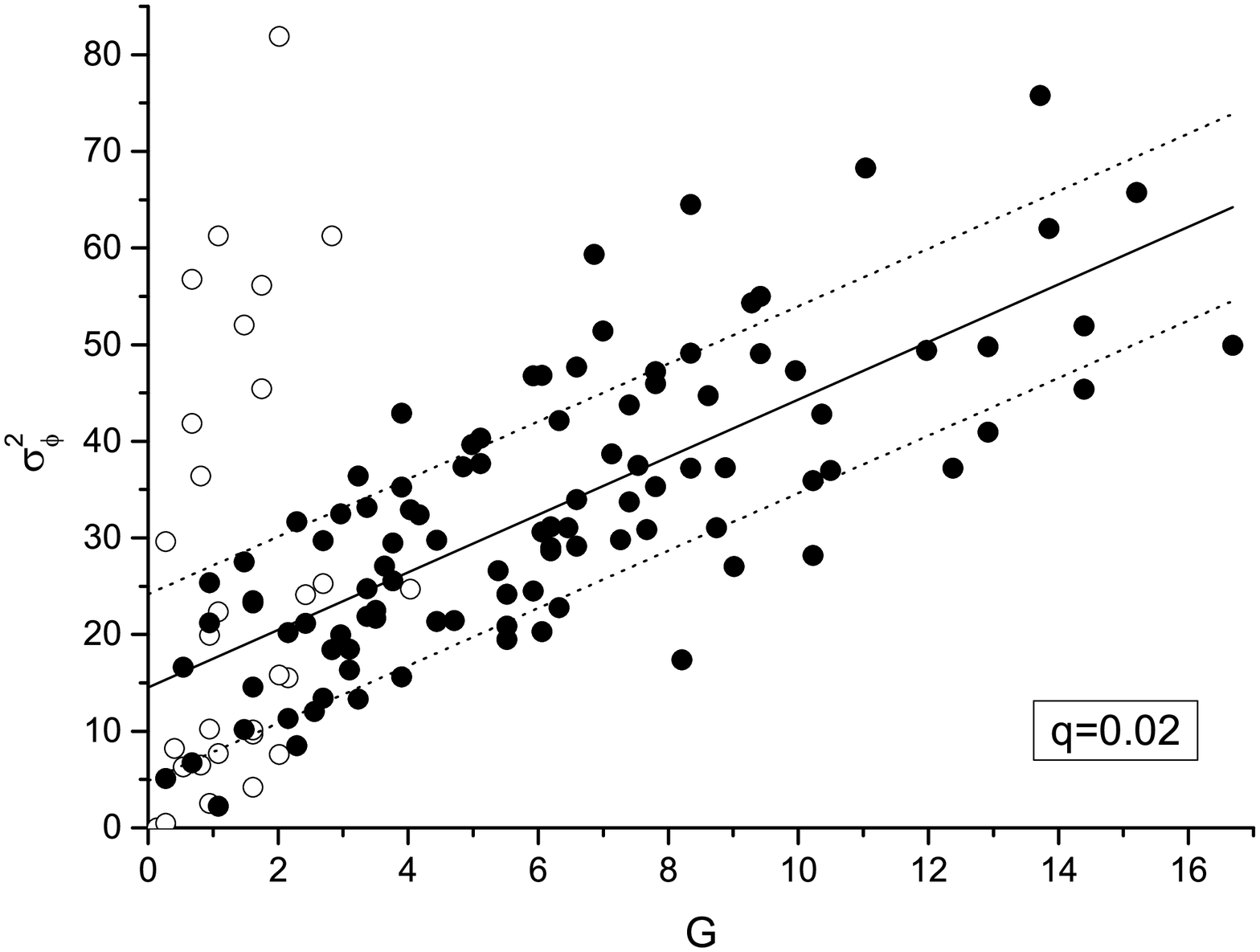}}
\caption{The regressions G --- $\sigma_\phi^2$ (\ref{reg}) for GC: $q=1$ (top) and $q=0.02$ (bottom).
The empty circles correspond to years of minimums and adjacent years,
which are not taken into account in building of the regressions.
The dash lines are for $a + b\,\mbox{G} \pm \Delta$,
where $\Delta$ is the rms of regression residuals (\ref{delta})} \label{fig2}
\end{center}
\end{figure}

In Fig.~\ref{fig2} the dependence G --- $\sigma_\phi^2$ for GC is
presented. We do not take into account years of cyclic minimums
and adjacent years (the empty circles in Fig.~\ref{fig1}),
because in that times "wings" of the ``Maunder butterflies'' tend to
overlap, and, therefore, $\sigma_\phi$ can be overestimated. The rest of data (the
filled circles) are described well (with the correlation
coefficient $r=0.88$) by the linear regression
\begin{equation}\label{reg}
\sigma_\phi^2 = a + b\,\mbox{G}\,,
\end{equation}
where $a = 13.6\pm1.0\;{\rm deg}^2$ and $b = 3.09\pm0.16\;{\rm deg}^2$.

\begin{figure}
\begin{center}
\ff{\includegraphics[width=0.75\textwidth]{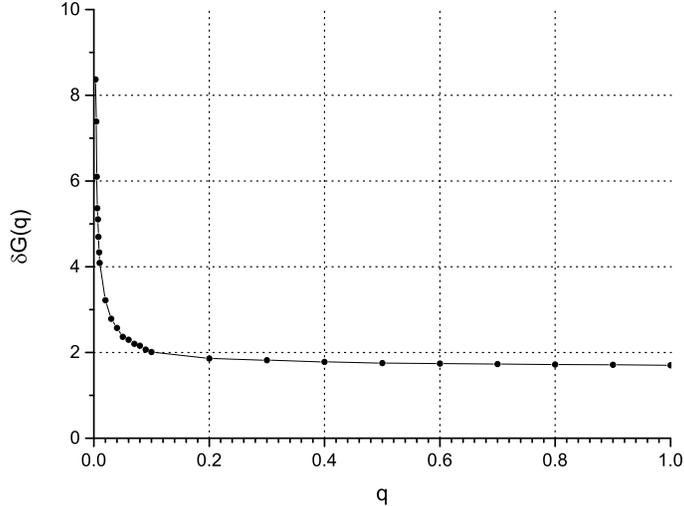}}
\caption{The dependence of the standard errors $\delta \mbox{G}$ on the loss ratio $q$ for GC.} \label{fig3}
\end{center}
\end{figure}

This relation is quite stable to loss of data. For example, if
to choose randomly only 2\% of sunspot groups observations from
GC
(hereafter we will refer to the ratio of the number of the residuary
observations of sunspot groups to their total number as
``the loss ratio'' $q$; in this case $q=0.02$), the errors raise, but the coefficients of the regression,
within the error limits, do not change: $a = 14.5\pm1.9\;{\rm deg}^2$
and $b = 2.98\pm0.26\;{\rm deg}^2$ ($r = 0.75$) (Fig.~\ref{fig2}b).
(The dependence of the coefficients on $q$ was in more details
discussed in \cite{im2016}.) One can use regression (\ref{reg}) to
obtain estimates of G by $\sigma_\phi^2$. The standard errors of
the obtained G can be estimated as $\delta \mbox{G} = \Delta/b$,
where
\begin{equation}\label{delta}
\Delta = \sqrt{\frac{1}{N}\sum_{i=1}^N \left((\sigma_\phi^2)_i - a - b\,\mbox{G}_i\right)^2}
\end{equation}
is the rms of the regression residuals (here the subscripts $i$ is
the number of the year for which indices are calculated). Strictly
speaking, one should calculate the residuals in (2) for a given
interval of G, but they are weakly dependent on G (see Fig.~\ref{fig2}) and we
can look for the estimate of errors summing over the total set of
indices $i$. The dependence of $\delta \mbox{G}$ on the loss ratio
$q$ is shown in Fig.~\ref{fig3}, where each point for the given $q$
was calculated as a mean of 12 random runs.

Having reconstructed G by known $\sigma_\phi$, we
can estimate the loss ratio $q$ as  $\mbox{G}_0/\mbox{G}$, where $\mbox{G}_0$
is (generally speaking, underestimated) ``index of sunspot groups'',
calculated by a fragmentary observational data.
After that we can find the error of reconstruction $\delta \mbox{G}$
for the given $q$ using the empirical dependence shown in Fig.~\ref{fig3}.

\begin{figure}
\begin{center}
\ff{\includegraphics[width=0.99\textwidth]{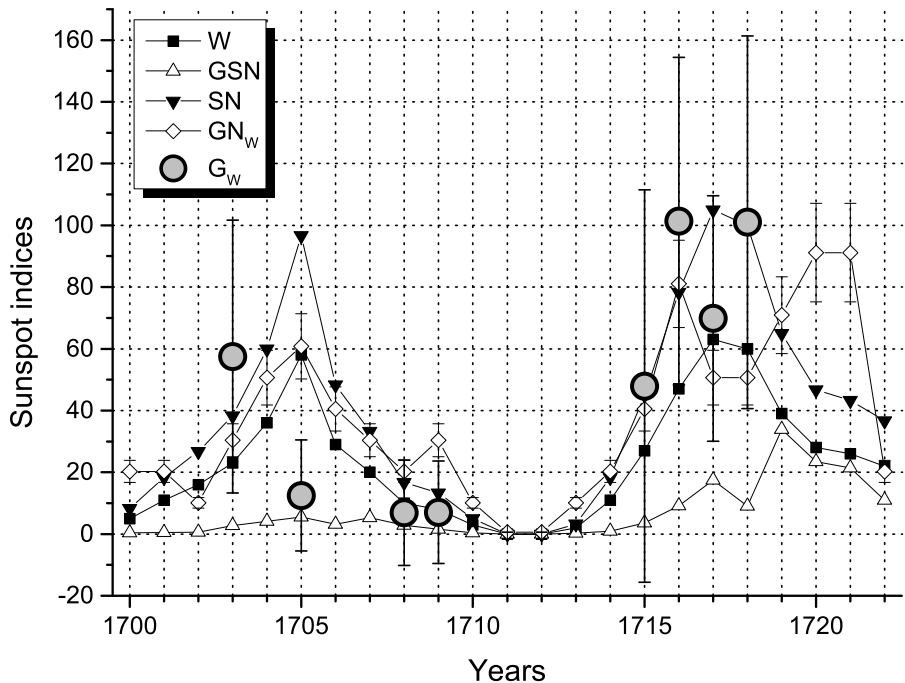}}
\caption{The sunspot indices (scaled to the Wolf number) in the end of the Maunder minimum.} \label{fig4}
\end{center}
\end{figure}

\section{Results and discussion}

Until the beginning of the 18th century the number of observation in
the catalog under investigation is too small to estimate the
latitude dispersion correctly. Therefore, we apply the
described method to the data starting from 1700. The estimates are
made only for years with four or more observations and
for the cases when it leads to positive values. Besides, in the first cycle
we have taken into account that sunspots existed in the south
hemisphere only and the estimate evaluated by the regression must be
divided by~2.

We will compare the estimates with other indices of activity known
for this epoch: W, GSN and their recently revised versions SN and GN
\cite{sn,gn,sngn}. To make the comparison more transparent it is
convenient to renormalize the obtained $\mbox{G}$, introducing
$\mbox{G}_{\mbox{w}} = 11.9\,\mbox{G}$, where the coefficient is
selected to minimize the rms difference between W and
$\mbox{G}_{\mbox{w}}$ for the epoch 1875--1976. The same procedure
is made for GN, leading to $\mbox{GN}_{\mbox{w}}=13.2\,\mbox{GN}$.

The correlation coefficient of yearly indices W,
$\mbox{G}_{\mbox{w}}$, GSN and $\mbox{GN}_{\mbox{w}}$ for the
Greenwich epoch 1875--1976 is higher than 0.98 and their rms
difference is less than 10 units. Therefore, for rough
estimates of activity in MM we do not make difference between these
four indices, expecting them to give approximately the same, by the
order of magnitude, level of activity.

In Fig.~\ref{fig4} we compare these indices and our estimates $\mbox{G}_{\mbox w}$ for years 1700--1719
(cycles Nos. $-3$ and $-4$ in Wolf's numeration). Comparison of
amplitudes and moments of maxima of cycles is made in
Table~\ref{tab1} (the asterisk marks the year of the first of two
maximums of $\mbox{GN}_{\mbox{w}}$ in cycle $-3$).

\begin{table}
\caption{Amplitudes and moments (in brackets) of
the 11-year cycles of solar activity Nos.~$-4$ and $-3$
in different sunspot indices.}\label{tab1}
\medskip
\small
\def\arraystretch{1.5}
\begin{tabular}{|c|l|l|l|l|l|}
\hline
No. & \multicolumn{5}{|c|}{Amplitudes of cycles and years of maximums}\\
\cline{2-6}
of cycle
& \multicolumn{1}{|c|}{$\mbox{G}_{\mbox{w}}$}
& \multicolumn{1}{|c|}{W}
& \multicolumn{1}{|c|}{GSN}
& \multicolumn{1}{|c|}{SN}
& \multicolumn{1}{|c|}{$\mbox{GN}_{\mbox{w}}$}
\\
\hline
$-4$ & $58\pm44$ (1703)  & 58 (1706) & 5.5 (1705) &  97 (1705)  & $70\pm12$ (1705) \\
$-3$ & $101\pm53$ (1716) & 63 (1717) & 34 (1719)  &  105 (1717) & $93\pm16$ (1716)* \\
\hline
\end{tabular}
\end{table}

\normalsize

One can see that our estimates of amplitudes are, in spite of large
uncertainties, in fair agreement with three indices (W, SN,
$\mbox{GN}_{\mbox{w}}$) and in significantly less agreement with
GSN. The latter index is lower for both cycles and its difference
from $\mbox{G}_{\mbox{w}}$ is more than 1.2 standard deviations; it
means that $\mbox{G}_{\mbox{w}}>\mbox{GSN}$ with probability
about 90\%. The moment of the sunspot latitude dispersion maximum in
cycle $-3$ also agrees with other data. For cycle $-4$ it is shifted
three years to the past, which can be a result of loss of data in
years 1704--1706.

Of course, the obtained estimates are correct under assumptions that a) the
latitudes of sunspots in the catalogs do not contain systematic
errors, and b) the linear regression (\ref{reg}) found for
``common'' epoch was the same in epochs of grand minimums. Under
these assumptions the latitude distribution of sunspots, in
agreement with the Wolf number and new revisited indices of activity
SN and GN, gives independent evidences in favor of not extremely low
cycles $-3$ and $-4$. Thus, the classical Wolf numbers, evidently,
describe solar activity in the end of the Maunder minimum more correct
than GSN do. The latitudinal data also confirm the conclusion (see
\cite{gn}) that the MM ended in the very beginning of the 18th
century rather than in 1720s.

\section{Acknowledgements}

The work was supported by the RFBR grant No.~16-02-00090
and the programs of the Presidium of the Russian Academy of Sciences
Nos.~21 and~22.

\end{document}